%% file: main.tex
\documentclass[12pt]{article}
\newcommand{\draft}{1} %
\input{definition}
\graphicspath{{./figs}}
\usepackage[symbol*]{footmisc}
\setfnsymbol{wiley}

\usepackage[final]{microtype}

\begin{document}

\begin{center}
{\bf\large Copula Structural Equation Models %
for Mediation Pathway Analysis}
\\\vskip0.3cm
  Canyi Chen, Ritoban Kundu, Wei Hao and Peter X.K. Song
\\\vskip0.3cm
\textit{Department of Biostatistics, University of Michigan}
\end{center}
\begin{abstract}
Structural equation models (SEMs) are fundamental to causal mediation pathway discovery. However, traditional SEM approaches often rely on \emph{ad hoc} model specifications when handling complex data structures such as mixed data types or non-normal data in which Gaussian assumptions for errors are rather restrictive. The invocation of copula dependence modeling methods to extend the classical linear SEMs mitigates several of key technical limitations, offering greater modeling flexibility to analyze non-Gaussian data. This paper presents a selective review of major developments in this area, highlighting recent advancements and their methodological implications.  
\end{abstract}

\section{Introduction}

We would like to congratulate Dr. Claudia Czado for her outstanding contributions to the field of vine copula models and their applications. Her exemplary career began at York University, Canada where Peter Song was lucky to be her colleague in the Department of Mathematics and Statistics, and later spectacularly flourished after she moved to the Technical University of Munich, Germany. We have had the privilege of witnessing her remarkable dedication and joy in advancing copula dependence models with a long list of publications appeared in various renowned journals. To celebrate Dr. Czado's amazing accomplishments, we want to dedicate this article with our great appreciation to one of her recent works \citep{czado2025VineCopulaBased}, which explores vine copula modeling approaches for studying graphical models, particularly directed acyclic graphs (DAGs). We hope to present our perspectives of copula dependence models for mediation analysis in connection to Dr. Czado's work in this important area. 

Pathway analysis is a cornerstone of numerous scientific disciplines, including biology, medicine, and sociology, as it facilitates the study of intricate interdependencies among variables under a hypothetical causal pathway. In epigenetics, for instance, pathway analysis is instrumental in identifying DNA methylation sites that mediate the effect of an exposure on an outcome of interest, thereby guiding potentially the development of targeted diagnostic and therapeutic strategies. These mediation relationships involve three primary variables, which are commonly structured and organized via directed acyclic graphs (DAGs) \citep{pearl2009CausalityModelsReasoning}. Consequently, the topology encodes the structural dependencies among variables. 
\par
Gaussian linear structural equation models (SEMs) provide a widely used framework for formulating
the likelihood and its decomposition under DAGs, enabling both estimation and inference of edge existence \citep{ullman2003StructuralEquationModeling}. Their applications span mediation analysis \citep{mackinnon2002ComparisonMethodsTest}, community detection \citep{zhang2022HighDimensionalGaussianGraphical}, and various other domains where conditional likelihoods are essential building blocks of the data generating mechanism. However, a fundamental limitation of Gaussian linear SEMs is the restrictive assumption of normality, which often limits the scope of data applications in practice. Empirical data are sampled frequently from skewness, heavy tails, or discrete distributions, necessitating more flexible modeling approaches that can accommodate complex, non-Gaussian dependencies and non-linear relationships in mediation analyses.

Copula-based models provide an appealing extension of the Gaussian data-generating mechanisms, significantly expanding the class of dependency structures to address complex data structures \citep{czado2019AnalyzingDependentData}. When integrated with SEMs, copula SEMs (CoSEMs) emerge as an useful alternative, offering greater modeling flexibility while preserving interpretability. \citet{elidan2010CopulaBayesianNetworks} explored the application of multivariate elliptical and Archimedean copulas in Bayesian networks, while \citet{hao2023ClassDirectedAcyclic} employed the Gaussian copula to introduce a hierarchical modeling approach for mediation analysis, later extended by \citet{chen2024QuantileMediationAnalytics} to investigate quantile mediation effects. More recently, \citet{czado2025VineCopulaBased} proposed a vine copula-based formulation of DAG modeling, where the inherent pairwise copula construction enables adaptive growth and pruning of the graph in terms of dimensionality and dependence complexity.   

Vine copulas \citep{joe2015DependenceModelingCopulas,czado2019AnalyzingDependentData,czado2022VineCopulaBased} have emerged as a powerful framework for modeling complex dependencies in multivariate data, providing a flexible alternative to traditional copula models such as Gaussian and Archimedean copulas. By decomposing a joint distribution into a sequence of bivariate copulas, vine copulas allow practitioners to capture intricate dependence structures while maintaining computational feasibility. This approach has been employed in applications across diverse fields, including finance, hydrology, and biomedical research, where understanding dependency patterns is essential. Despite their advantages, vine copula decompositions introduce interpretability challenges, particularly in applications where causal topological structures encoded in DAGs must be preserved.  

Dr. Claudia Czado's recent work \citep{czado2025VineCopulaBased}, provides a new vine copula based formulation for DAGs and examines theoretical properties and practical applications. This works extends the applicability of vine copulas beyond conventional Gaussian SEMs by incorporating non-linear dependencies. Traditional Gaussian linear SEMs, commonly used in DAG-based Bayesian networks, may fail to capture the complex non-linear dependencies prevalent in real-world data, limiting their ability to model nonlinear, non-additive, and non-Gaussian conditional distributions. To address these limitations, Czado proposes a D-vine copula-based regression framework to specify the conditional distribution of an offspring node given its parents. Her proposed approach substantially generalizes the standard linear regression paradigm by accommodating flexible dependence structures along with the use of various non-linear regression models. Moreover, it facilitates the ranking of parent nodes by their relative importance and provides a principled mechanism for pruning edges in an initial DAG structure when empirical evidence does not support the existence of certain pathways. In addition, the D-vine copula-based SEM facilitates uncertainty quantification and demonstrates substantial improvements over Gaussian linear SEMs.

A major challenge associated with vine copula decompositions is the \textit{non-uniqueness} of the construction, which can lead to complications in both data analysis and interpretation. In contrast, probabilistic graphical models impose a well-defined structure on conditional dependencies which vine copulas allow multiple equivalent decompositions of the same joint distribution. The flexibility of vine-copula modeling permits different conditional distribution representations of the same data, potentially influencing the conclusions drawn from the analysis. Consequently, ensuring consistency and interpretability remains a key challenge when applying vine copulas to causal inference and pathway analysis in which the identifiability conditions are required.

Preserving topological ordering or \textit{acyclicity} is particularly critical in settings where hypothesized pathways must be maintained. DAGs serve as foundational tools for representing directed relationships which are the basis employed in pathway analyses. However, the current formulation of vine copula decompositions do not inherently enforce acyclicity, potentially introducing cyclic dependencies that obscure causal interpretations. Ensuring that statistical models adhere to the underlying topological ordering is essential for drawing meaningful inferences and maintaining interpretability in mediation analysis. In this regard, the Gaussian copula SEM proposed by \citet{hao2023ClassDirectedAcyclic} is particularly noteworthy, as it accommodates non-Gaussian data while preserving the DAG acyclicity structure and ensuring interpretability.
\par
The remainder of the paper is organized as follows. Section \ref{section:gaussian_SEM} revisits classical Gaussian SEMs and reformulates the corresponding DAG topological 
structure, laying the foundation for hierarchical modeling. Section \ref{section:copula_SEM} reviews copula SEMs, including both Gaussian copula SEMs and vine copula SEMs. Section \ref{section:ab_test} introduces the adaptive bootstrap method for hypothesis testing under a composite null in mediation pathway analysis. In Section \ref{section:numerical_studies}, we present a numerical study demonstrating the effectiveness of the adaptive bootstrap method compared to the widely used vanilla bootstrap approach. Finally, Section \ref{section:conclusion} concludes the paper and discusses potential directions for future research.  

\section{Classical Structural Equation Model}\label{section:gaussian_SEM}
\begin{figure}[tbph!]
\centering{\renewcommand{\arraystretch}{0.8} %
\begin{tabular}{c}    
$$
\begin{xy}
\xymatrix{
&&&\\
\W \ar[r] \ar@/^1.5pc/[rr] \ar@/^2.5pc/[rrr] & 
S \ar[r]^{\alpha} \ar@/_1.5pc/[rr]_{\beta} & 
M \ar[r]^{\gamma} & 
Y
}
\end{xy}
$$
\end{tabular}
}
\captionsetup{font = footnotesize}
\caption{
A directed acyclic graph with topologically ordered scalar exposure $S$, scalar mediator $M$, scalar outcome $Y$, and $p$-dimensional confounders $\W$, where $\alpha, \beta, \gamma$ are DAG coefficients for the relationships.
}\label{fig:mediationDAG_unconfounded}
\end{figure}
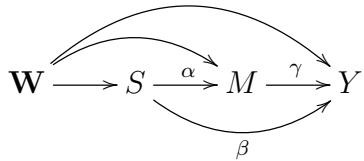

In this section, we revisit the classical Gaussian SEMs in mediation pathway analysis. Mediation analysis seeks to determine whether an intermediate variable mediates the effect of environmental or social exposures on an outcome of interest, given a hypothesized prespecified directed acyclic graph (DAG); see, for example, \cite{sobel1982AsymptoticConfidenceIntervals} and \cite{baron1986ModeratorMediatorVariable}. A single DAG,  depicted in \Cref{fig:mediationDAG_unconfounded},  includes a scalar exposure $S$, a scalar mediator $M$, a scalar outcome $Y$ and a $p$-dimensional vector of confounders $\W$. The coefficients $\alpha$, $\beta$, and $\gamma$ characterize the relationships among these variables.  

To model the DAG shown in \Cref{fig:mediationDAG_unconfounded}, the classical Gaussian SEM consists of the following three linear equations: 
\begin{align*}
S &= \mathbf{W} \boldsymbol{\beta}_x +\epsilon_x;\\ \notag
M & = \alpha S +  \mathbf{W} \boldsymbol{\beta}_m +\epsilon_m;  \notag   \\  
 	Y  & = \gamma S +  \beta  M + \mathbf{W} \boldsymbol{\beta}_y +  \epsilon_y  \notag
\end{align*}
where $(\epsilon_x, \epsilon_m, \epsilon_y)^\top \sim \calN(\bzeros, \bSig)$ with $\bSig = \diag(\sigma_x^2, \sigma_m^2, \sigma_y^2)$. 
The acyclicity in the DAG is fully described by the lower-triangular matrix $\bTheta$ in the joint distribution of the form:  
\begin{align*}
\left( \begin{array}{c} S \\  M \\  Y \end{array} \right) 
  & = 
\bTheta \left( \begin{array}{c} S \\ M \\ Y \end{array} \right) + 
 \left( \begin{array}{c}  \mathbf{W} \boldsymbol{\beta}_s \\  \mathbf{W} \boldsymbol{\beta}_m \\  \mathbf{W} \boldsymbol{\beta}_y \end{array} \right)  + \left( \begin{array}{c}  \epsilon_s \\ \epsilon_m \\ \epsilon_y \end{array} \right),\text{  } \bTheta\defn \left(\begin{array}{ccc} 0 & 0 & 0 \\  \alpha & 0 & 0 \\ \gamma & \beta & 0 \end{array}\right),
 \end{align*} 
which can be rewritten as follows:
$$
(\eye-\bTheta) \left( \begin{array}{c} S \\  M \\  Y \end{array} \right) 
= \left( \begin{array}{c}  \mathbf{W} \boldsymbol{\beta}_s \\  \mathbf{W} \boldsymbol{\beta}_m \\  \mathbf{W} \boldsymbol{\beta}_y \end{array} \right)  + \left( \begin{array}{c}  \epsilon_x \\ \epsilon_m \\ \epsilon_y \end{array} \right). 
$$

This vector-based expression conveys two key insights. (i) The confounding effects $\mathbf{W}$ influence the SEM solely through the marginal means (i.e., the first moments) of the errors $(\epsilon_x, \epsilon_m, \epsilon_y)^\top$; they do not alter the covariance structure of the joint distribution of the errors. (ii) The acyclic topology of the DAG is encoded in the lower triangular matrix $\bTheta$, which appears in the SEM through the covariance (i.e., the second moments). Specifically, the resulting covariance takes the form:
\begin{eqnarray}
    \label{eq: cov}\bGam \defn \cov\{(S, M, Y)\trans; \mathbf{W}\} = (\eye-\bTheta)^{-1} \bSig (\eye-\bTheta)^{-\top},
\end{eqnarray}
which remains independent of the confounders $\W$. The DAG parameters $(\alpha, \beta, \gamma)$ do not affect the marginal means of the joint distribution of the errors. Such separability brings desired convenience in statistical analysis and interpretability. \citet{ZhouBCS2021} proposed a generalization of $\bSig$ from a diagonal matrix to a low-rank latent factor model in the study of network topology with directed edges.

\section{Copula Structural Equation Model}\label{section:copula_SEM}
This section reviews two copula SEMs that significantly extend the classical Gaussian SEM to accommodate non-Gaussian data.
\subsection{Gaussian Copula SEMs}\label{subsec:gaussian_copula}
\cite{hao2023ClassDirectedAcyclic} proposed leveraging Gaussian copulas to fully exploit hierarchical structures while preserving the DAG topological structure. Their approach, termed {\it generalized structural equation models} (GSEMs), is primarily designed to accommodate mixed data types.   

As established in Section \ref{section:gaussian_SEM}, the DAG structure in mediation analysis is encoded by the lower triangular matrix $\bTheta$, which induces the covariance matrix  $\bGam = (\eye-\bTheta)^{-1} \bSig (\eye-\bTheta)^{-\top}$; see equation \eqref{eq: cov}.  
Building on this line of thought, \cite{hao2023ClassDirectedAcyclic} incorporated the covariance specification of $\bGam$ in \eqref{eq: cov}  into a valid joint probability distribution for $S$, $M$, and $Y$, thereby extending the framework to nonlinear, non-normal GSEMs.  According to the insights gained from the classical SEMs, confounders do not appear in the covariance matrix $\bGam$ but instead influence the marginal distributions of individual variables. This hierarchical formulation effectively disentangles marginal and dependence parameters within the GSEM framework. Notably, the proposed GSEMs encompass classical linear normal SEMs as a special case when the errors are normally distributed.

By Sklar’s Theorem \citep{sklar1959fonctions}, if $(S, M, Y)$ are continuous random variables, there exists a copula $C$ such that 
\[
F(s, m, y) = C\{F_s(s), F_m(m), F_y(y); \wt \bGam\},
\]  
where $F_j(\cdot)$ and $f_j(\cdot)$ denote the cumulative distribution function and density function, respectively. Here the copula function $C(\cdot)$ is independent of the marginal parameters.  To embed the DAG covariance structure in (\ref{eq: cov}) within the copula framework, \cite{hao2023ClassDirectedAcyclic} employ the Gaussian copula \citep{xue2000multivariate} of the following form  
\begin{equation}\label{eq: Gaussian}
C(u_1, u_2, u_3) = \Phi_3\{\Phi^{-1}(u_1),\Phi^{-1}(u_2),\Phi^{-1}(u_3); \wt \bGam\},
\end{equation}  
where $\Phi_3(\cdot ; \wt\bGam)$ denotes the trivariate Gaussian distribution function with mean zero and correlation matrix $\wt\bGam$, $\Phi(\cdot)$ represents the standard univariate Gaussian distribution function, and $\Phi^{-1}(\cdot)$ is its normal quantile function. 
Here $\wt \bGam$ may be derived from $\bGam$ in \eqref{eq: cov} by scaling the diagonal elements to be all equal to $1$. 
Since all marginal parameters, including variances, are confined within the marginal distributions, the dependence matrix $\wt\bGam$ consists solely of correlation parameters, independent of marginal variances. {\color{black}As a result, the parameters in $\wt\bGam$ quantify nonlinear measures of pairwise associations, which are closely related to rank-based correlations such as Kendall's Tau and Spearman's Rho, allowing the model to capture a broader range of dependencies beyond the standard Pearson correlations.   }
 
The construction of GSEMs is based on a joint distribution induced by the Gaussian copula model in (\ref{eq: Gaussian}), which accommodates both continuous and discrete marginal distributions while incorporating a structured dependence matrix $\wt\bGam$ derived from  (\ref{eq: cov}). As shown in \citep{hao2023ClassDirectedAcyclic}, this framework enables the modeling of mixed data types, making it particularly suitable for mediation analyses.  We discuss two distinct settings: GSEMs without confounders and GSEMs accounting for confounders in observational studies.

 In the first, each marginal distribution follows a generalized linear model (GLM). Specifically, the variables $S$, $M$, and $Y$ are, respectively, modeled as members of the exponential dispersion (ED) family \citep{jorgensen1987exponential}:  
\[
S \sim ED(\mu_s,\phi_s), \quad M \sim ED(\mu_m,\phi_m), \text{ and } Y \sim ED(\mu_y,\phi_y),
\]
where $(\mu_s, \mu_m, \mu_y)$ are the mean parameters, and $(\phi_s, \phi_m, \phi_y)$ are the dispersion parameters, all independent of covariates. Setting $u_1 = F_s(s)$, $u_2 = F_m(m)$, and $u_3 = F_y(y)$  yields a  joint distribution  
$
\pi(s, m, y; \wt\bGam) = C(u_1, u_2, u_3; \wt\bGam)$,  
with $C(\cdot)$ being the Gaussian copula function in (\ref{eq: Gaussian}).  

It is interesting to note that under the assumptions in (\ref{eq: cov}) and (\ref{eq: Gaussian}), the resulting joint distribution $\pi(s, m, y; \wt\bGam)$ can be equivalently expressed through a latent variable representation, through which an organic connection to the classical SEM may be established. Specifically, let $Z_s$, $Z_m$, and $Z_y$ be latent normal random variables satisfying  
\begin{eqnarray}\label{latent_representation}
Z_s \sim N(0, 1),  \enskip
Z_m \mid Z_s \sim N(\alpha Z_s, 1), \text{ and }
Z_y \mid Z_s, Z_m \sim N(\gamma Z_s + \beta Z_m, 1). \label{eq: general_model}
\end{eqnarray}  
This formulation \eqref{latent_representation} provides a direct interpretation connecting to the DAG dependence structure within GSEMs, reinforcing their flexibility in capturing complex relationships while preserving the topological ordering among continuous, discrete, or mixed data types. 

The mapping between the observed variables $(S,M,Y)$ and the latent variables $(Z_s, Z_m, Z_y)$ can be easily  established through the monotonic marginal quantile transformations:  
$
S= F_s^{-1}\{\Phi(Z_s)\}$, $M= F_m^{-1}\{\Phi(Z^*_m)\}$, and $Y= F_y^{-1}\{\Phi(Z^*_y)\},
$
where  
$
Z_m^* = {Z_m}/{\tau_m}$, $Z_y^* = {Z_y}/{\tau_y}$,  $\tau_m = \sqrt{\alpha^2+1}$, and $ \tau_y = \sqrt{(\gamma+\alpha\beta)^2+\beta^2+1}$ with the scaling operations to ensure the unitary diagonal elements in $\wt \bGam$.  
As a result, the marginal distributions of $Z_s$, $Z_m^*$, and $Z_y^*$ follow the standard normal distribution, while their joint distribution is a trivariate normal.

If $S$ is continuous, the transformation simplifies to $Z_s = \Phi^{-1}(F_s(s))$, the quantile of the standard normal distribution. If $S$ is discrete, taking values in $\{0,1,2,\dots\}$, then  $S = \sum_{s=0}^{\infty} s I(F_s(s-1) \leq \Phi(Z_s) < F_s(s))$. 
This implies that the event $\{S = s\}$ is equivalent to $\{\Phi^{-1}(F_s(s-1)) \leq Z_s < \Phi^{-1}(F_s(s))\}$. Similar arguments are applied to $M$ and $Y$ via $Z_m^*$ and $Z_y^*$. %

To account for confounders, \cite{hao2023ClassDirectedAcyclic} incorporate them into the univariate GLMs specifying the marginal distributions:  $S \mid \bfW \sim ED(\mu_s,\phi_s)$ with $g_s(\mu_s) = \bfW^\top \bfbeta_s$, $M \mid \bfW \sim ED(\mu_m,\phi_m)$ with  $g_m(\mu_m) = \bfW^\top \bfbeta_m$, and $Y \mid \bfW \sim ED(\mu_y,\phi_y)$ with $g_y(\mu_y) = \bfW^\top \bfbeta_y$, 
where $\bfbeta_s$, $\bfbeta_m$, and $\bfbeta_y$ are regression coefficient vectors, and $g_s$, $g_m$, and $g_y$ are suitable link functions. This formulation ensures that confounders influence the marginal distributions while preserving the DAG topological dependence structure in the copula.

\subsection{Vine Copula SEMs}\label{section:vine_copula}

To relax the use of Gaussian copulas in Section~\ref{subsec:gaussian_copula}, \cite{czado2025VineCopulaBased} proposed the use of vine copulas to capture multivariate dependencies by decomposing the joint distribution into bivariate copulas and marginal distributions. In our case of three variables, the vine copula formulation suggests that the joint density of $S,M,Y$, $f(s, m, y)$  can be written as, with $x_1=s,x_2=m$ and $x_3=y$,  
\begin{eqnarray*}
     c_{13;2}(F_{1|2}(x_1|x_2), F_{3|2}(x_3|x_2)) \times c_{23}(F_2(x_2), F_3(x_3)) \times c_{12}(F_1(x_1), F_2(x_2)) {\color{black}\times}  \prod_{j=1}^3 f_j(x_j).
\end{eqnarray*}
Here, \( c_{13;2} \) denotes the conditional copula density of \( (S, Y) \) given \( M \), while \( c_{12} \) and \( c_{23} \) are unconditional bivariate copulas. Vine copulas are typically constructed using hierarchical tree structures, such as D-vines (sequential path structures) and C-vines (star-like structures). These models are widely applied in finance, risk management, and Bayesian networks. The R package \texttt{rvinecopulib} \citep{nagler2017RvinecopulibHighPerformance} provides efficient numerical tools for estimating simplified vine copula models, balancing flexibility with computational efficiency.  
\par
Vine copulas can also be employed in regression models, enabling the capture of nonlinear relationships by modeling the conditional distribution of the response given covariates. Two principal approaches exist in the literature. The {\it covariates-first} approach constructs a vine model for the covariates and subsequently incorporates the response \citep{chang2019PredictionBasedConditional,pan2022PredictingTimesEvent}. Conversely, the {\it response-first} approach begins with the response variable and sequentially adds covariates \citep{kraus2017DvineCopulaBased,zhu2021SimplifiedRvineBased}. This provides a perspective from which \citet{czado2025VineCopulaBased} constructs the vine copula SEMs.

Consider a generic setting with response variable \( Y \) and covariates \( X_1, \dots, X_k \). The D-vine copula regression first transforms these variables to the copula scale:
\[
V = F_Y(Y), \quad U_j = F_{X_j}(X_j), \quad j = 1, \dots, k.
\]
On this scale, the joint copula density takes the form
\[
c_{V,1,\dots,k}(v, u_1, \dots, u_k) = \left[ \prod_{j=1}^k c_{V,j;1,\dots,j-1} \right] \left[ \prod_{j=1}^{k-1} \prod_{i=1}^{k-j} c_{i,(i+j);(i+1),\dots,(i+j-1)} \right],
\]
where \( c_{V,j;1,\dots,j-1} \) denotes the copula density of \( (Y, X_j) \) given \( X_1, \dots, X_{j-1} \). The conditional density of \( V \) given \( U_1 = u_1, \dots, U_k = u_k \) is then given by
\[
c_{V|U_1,\dots,U_k}(v|u_1, \dots, u_k) = \prod_{j=1}^k c_{V,j;1,\dots,j-1}.
\]
The above formulation facilitates to yield the conditional quantile function  
\[
F^{-1}_{Y|X_1,\dots,X_k}(\alpha|\mathbf{x}) = F^{-1}_Y \left( C^{-1}_{V|U_1,\dots,U_k}(\alpha|F_1(x_1), \dots, F_k(x_k)) \right).
\]
The technique above is utilized by \cite{czado2025VineCopulaBased} to develop D-vine copula structural equation models (SEMs). Extending the classical Gaussian DAG framework by incorporating D-vine copulas, allows for nonlinear dependencies between parent and child nodes. For a node \( X_i \) with parents \( \boldsymbol{\pi}(X_i) = (X_{j_1(i)}, \dots, X_{j_{n_i}(i)}) \), the conditional density takes the form
\begin{equation}
    f(x_i|\boldsymbol{\pi}(X_i) = \boldsymbol{\pi}(x_i)) = \left[ \prod_{s=2}^{n_i} c_{i,s;1:s-1} \right] \times c_{i,1} \times f_i(x_i), \label{eq:vine}
\end{equation}
where the conditional copula density of \( (X_i, X_s) \) given \( X_{j_1(i)}, \dots, X_{j_{s-1}(i)} \) is given by,
\[
c_{i,s;1:s-1} = c_{i,s;1:s-1}(F_{i|1:s-1}(x_i|x_{j_1(i)}, \dots, x_{j_{s-1}(i)}), F_{s|1:s-1}(x_s|x_{j_1(i)}, \dots, x_{j_{s-1}(i)}))
\]
and the marginal copula density of \( (X_i, X_{j_1(i)}) \) is
\[
c_{i,1} = c_{i,1}(F(x_i), F(x_{j_1(i)})).
\]
This construction in equation \eqref{eq:vine} serves as a fundamental component for factorizing the joint distribution within the SEM, ensuring consistency with the underlying conditional structure due to topological ordering among variables in the DAG.  
\par
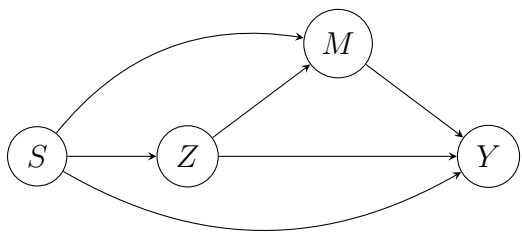
\begin{figure}[htbp!]
\begin{center}
\begin{tikzpicture}[>=stealth, node distance=2cm]
    \node (S) at (0,0) [circle, draw] {$S$};
    \node (Z) at (2,0) [circle, draw] {$Z$};
    \node (M) at (4,1.5) [circle, draw] {$M$};
    \node (Y) at (6,0) [circle, draw] {$Y$};

    \draw[->] (S) -- node[above] {} 
    (Z);
    \draw[->] (Z) to node[sloped, above] {} (M);
    \draw[->] (M) to node[sloped, above] {} (Y);
    \draw[->, bend right] (S) to node[below] {} (Y);
    \draw[->, bend left] (S) to node[above] {} (M);
    \draw[->] (Z) to node[below] {} (Y);
\end{tikzpicture}
\end{center}
\caption{Directed acyclic graph (DAG) with exposure $S$ that influences the outcome $Y$ through two sequentially ordered mediators, $Z$ and $M$.}
\label{fig:mediationDAG}
\end{figure}

This methodology is particularly well-suited for modeling sequentially ordered mediators in mediation analysis where multiple mediators arise, \emph{say},  in a temporal order.  Consider a DAG with two sequentially ordered mediators $Z$ and $M$ in Figure \ref{fig:mediationDAG}.  
For the random vector \( (S, Z, M, Y) \), the joint density \( f_{S,Z,M,Y} \) is decomposed using a D-vine structure as follows:
\beqr
\notag f_{S,Z,M,Y}(x_S, x_Z, x_M, x_Y) &= & \, c_{SY|ZM}\left(F_{S|ZM}(x_S|x_Z, x_M), F_{Y|ZM}(x_Y|x_Z, x_M)\right) \\
\notag& &\times c_{SM|Z}\left(F_{S|Z}(x_S|x_Z), F_{M|Z}(x_M|x_Z)\right) \\
\notag&& \times c_{ZY|M}\left(F_{Z|M}(x_Z|x_M), F_{Y|M}(x_Y|x_M)\right) \\
\notag&& \times c_{SZ}\left(F_S(x_S), F_Z(x_Z)\right) \\
\notag&& \times c_{ZM}\left(F_Z(x_Z), F_M(x_M)\right) \\
\notag&& \times c_{MY}\left(F_M(x_M), F_Y(x_Y)\right) \\
&& \times f_Y(x_Y) f_M(x_M) f_Z(x_Z) f_S(x_S).\label{equation:vine_decomposition}
\eeqr
Each term in the decomposition corresponds to a specific pathway in the DAG. The conditional dependence between \( S \) and \( Y \) given \( Z \) and \( M \) is captured by \( c_{SY|ZM} \), while \( c_{SM|Z} \) represents the dependence between \( S \) and \( M \) given \( Z \), and \( c_{ZY|M} \) accounts for the dependence between \( Z \) and \( Y \) given \( M \). In addition, the terms \( c_{SZ}, c_{ZM}, \) and \( c_{MY} \) represent marginal dependencies, while \( f_Y, f_M, f_Z, \) and \( f_S \) denote the corresponding marginal density functions. {\color{black}In the given vine copula expression (\ref{equation:vine_decomposition}), based on the discussion in Section~\ref{section:gaussian_SEM} on the construct of classical SEMs, it is reasonable to assume that covariates do not influence the conditional copula but only the marginals. Technically, this assumption is less restrictive than conditional independence according to \cite{czado2022VineCopulaBased}; rather, it is organically imposed for the modeling of mediation effects under the DAG for interpretability.}\\

\noindent
{\color{black} For the DAG in Figure~\ref{fig:mediationDAG}, \citet{zhou2022semiparametric} identifies three primary causal pathways to decompose the total effect of $S$ on $Y$, including (i) the direct path $S \rightarrow Y$; (ii) the indirect path via mediator $M$, i.e., $S \rightarrow M \rightarrow Y$; and (iii) a composite path $S \rightarrow Z \leadsto Y$, where path $Z \leadsto Y$ encompasses a collection of all pathways from $Z$ to $Y$ comprised of both $Z \rightarrow Y$ and $Z \rightarrow M \rightarrow Y$. The latter two paths involving $Z$ are collapsed for a joint mass because they are individually identifiable under general regularity conditions.
Under the structure of the decomposition in  \eqref{equation:vine_decomposition}, we can perform hypothesis testing for the conditional independence to investigate the presence or absence of specific pathways. A few scenarios of interest are listed below. 
\begin{itemize}
    \item To test for the presence of a direct effect $S \rightarrow Y$, conditional on $Z$ and $M$, we examine whether the conditional copula is the independence copula: 
$$
c_{SY|ZM}\left(F_{S|ZM}(x_S|x_Z, x_M), F_{Y|ZM}(x_Y|x_Z, x_M)\right) = F_{S|ZM}(x_S|x_Z, x_M) \times F_{Y|ZM}(x_Y|x_Z, x_M).
$$
If it holds, we claim that the absence of a direct effect from $S$ to $Y$ after adjusting for $Z$ and $M$.
\item To test for the mediated pathway $S \rightarrow M \rightarrow Y$, we jointly assess the null hypotheses: independence between $S$ and $M$ given $Z$ and/or independence between $M$ and $Y$. That is, these two scenarios of independence take the following form: 
\begin{eqnarray*}
c_{SM|Z}\left(F_{S|Z}(x_S|x_Z), F_{M|Z}(x_M|x_Z)\right) & = & F_{S|Z}(x_S|x_Z) \times F_{M|Z}(x_M|x_Z), \\
 c_{MY}\left(F_M(x_M), F_Y(x_Y)\right) & =  & F_M(x_M)\times F_Y(x_Y).
\end{eqnarray*}
If either holds, then this hypothesized mediation pathway $S \rightarrow M \rightarrow Y$ is deemed absent.
\item  Testing the composite pathway $S \rightarrow Z \leadsto Y$ is equivalent to testing whether either pair copula below is the independence copula, namely
\begin{eqnarray*}
c_{SZ}\left(F_S(x_S), F_Z(x_Z)\right) &= & F_S(x_S)\times F_Z(x_Z) \\
c_{ZY|M}\left(F_{Z|M}(x_Z|x_M), F_{Y|M}(x_Y|x_M)\right) & = & F_{Z|M}(x_Z|x_M) \times F_{Y|M}(x_Y|x_M). 
\end{eqnarray*}
\end{itemize}
There are many approaches to testing for the independence copula. To our own knowledge, we suggest the means of Shannon's mutual information. It is known that independence may be equivalently assessed by zero mutual information. This assessment can be performed effectively using a recent nonparametric test method proposed in  \citet{purkayastha2023quantification}, which is shown to enjoy both computational efficiency and statistical consistency.}

\section{Statistical Analysis}
\subsection{Estimation}
Once a copula SEM is specified, one primary task pertains to parameter estimation. Given that the generalized SEM is fully parametric, the maximum likelihood estimation (MLE) is a natural choice of the method for parameter estimation.  In the implementation of MLE, a direct solution with all modeling parameters may be carried out with existing optimization recipes such as R function \textsf{optim()}. Alternatively, one can adopt the widely used strategy proposed by \cite{joe2005AsymptoticEfficiencyTwostage} in the literature of copula dependence models. The so-called Inference Function with Marginals (IFM) is known to be computationally efficient in handling nuisance parameters in the marginals.  For example, in the Gaussian copula SEMs proposed by \cite{hao2023ClassDirectedAcyclic}, given a random sample $\{(S_i, M_i, Y_i, \W_i)\}_{i = 1}^n$ of size $n$, one may run MLE to obtain the estimates of the model parameters $(\wh\alpha, \wh\beta, \wh\gamma, \wh\bbeta_s\trans, \wh\bbeta_m\trans, \wh\bbeta_y\trans, \wh\phi_s, \wh\phi_m, \wh\phi_y)$ where the DAG parameter estimates  $(\wh\alpha, \wh\beta, \wh\gamma)$ are calculated after the marginal model parameter estimates $(\wh\bbeta_s\trans, \wh\bbeta_m\trans, \wh\bbeta_y\trans, \wh\phi_s, \wh\phi_m, \wh\phi_y)$ are obtained. As shown by \cite{joe2005AsymptoticEfficiencyTwostage}, IFM produces asymptotically efficient MLE.

\subsection{Hypothesis Testing}\label{section:ab_test}

Another task is to test for the presence of a mediation effect or the existence of a pathway. {\color{black}As shown by \cite{hao2023ClassDirectedAcyclic}, within the Gaussian copula SEM framework described in Section~\ref{subsec:gaussian_copula}, the mediation effect of $M$  on the association of exposure $S$ and the mean of  $Y$ is proportional to the product $\alpha\beta$. Extending this result, \cite{chen2024QuantileMediationAnalytics} re-established this proportionality at a quantile level of $Y$.
 }  Given the DAG  parameters $\alpha$ and $\beta$, {\color{black}testing the pathway effect $S \to M \to Y$} is typically framed as the hypothesis test of the following null and alternative hypotheses:  
\[
H_0\colon \alpha\beta = 0 \quad \text{versus} \quad H_1\colon \alpha\beta\neq 0.
\]  
This testing problem is challenging due to the irregular behavior of the sample estimator $\wh\alpha\wh\beta$ under different regions of the null. Specifically, the issue of singularity arises at $(\alpha, \beta) = (0,0)$, where the gradient of $\alpha\beta$ with respect to $\alpha$ and $\beta$ becomes zero.  

To implement the Sobel test via {\color{black}an} adaptive bootstrap, we begin by assuming  $n^{1/2}(\wh\alpha - \alpha)\todistribution\calN(0, \sigma_{\alpha}^2)\defn Z_\alpha$ and $n^{1/2}(\wh\beta - \beta)\todistribution\calN(0, \sigma_{\beta}^2)\defn Z_{\beta}$. According to 
\citet{he2023AdaptiveBootstrapTests}, we have the following large sample properties: (i) when $\left(\alpha, \beta\right) \neq(0,0), \sqrt{n} (\wh{\alpha} \wh{\beta}-\alpha \beta) \xrightarrow{d} \alpha Z_{\beta}+\beta Z_{\alpha}$; and (ii) when $\left(\alpha, \beta\right)=(0,0), n \times(\wh{\alpha} \wh{\beta}-\alpha \beta) \xrightarrow{d}Z_{\alpha} Z_{\beta}$, as $n\to\infty$.

We introduce the adaptive bootstrap (AB) approach proposed by \cite{he2023AdaptiveBootstrapTests}, which allows us to address the singularity issue described above. To distinguish three null cases, namely $\{\alpha = 0, \beta \neq 0\}$, $\{\alpha\neq 0, \beta = 0\}$ and $\{\alpha = 0, \beta = 0\}$, a decomposition of the test statistic is provided to isolate the scenario of $\left(\alpha, \beta\right) = (0,0)$ from other two null cases. This is achieved by evaluating the absolute values of the standardized statistics $T_{\alpha} = \sqrt{n} \wh\alpha / \hat{\sigma}_{\alpha}$ and $T_{\beta} = \sqrt{n} \wh\beta / \hat{\sigma}_{\beta}$ against a predefined threshold, where $\hat{\sigma}_{\alpha}$ and $\hat{\sigma}_{\beta}$ denote the sample standard deviations of $\wh\alpha$ and $\wh\beta$, respectively. Specifically, \cite{he2023AdaptiveBootstrapTests} established the  following decomposition  
\begin{eqnarray}\label{equation:decomposition}
    \wh\alpha \wh\beta-\alpha \beta=  \left(\wh\alpha \wh\beta-\alpha \beta\right) \times\left(1-\mathrm{I}_{\alpha, \lambda_n} \mathrm{I}_{\beta, \lambda_n}\right)  +\left(\wh\alpha \wh\beta-\alpha \beta\right) \times \mathrm{I}_{\alpha, \lambda_n} \mathrm{I}_{\beta, \lambda_n},
\end{eqnarray}
with two indicator functions  $\mathrm{I}_{\alpha, \lambda_n} = \mathrm{I}\left\{\left|T_{\alpha}\right| \leq \lambda_n, \alpha=0\right\}$ and $\mathrm{I}_{\beta, \lambda_n} = \mathrm{I}\left\{\left|T_{\beta}\right| \leq \lambda_n, \beta=0\right\}$. Here $\mathrm{I}\{E\}$ represents the indicator function of an event $E$, and $\lambda_n$ denotes a specified threshold. In either null $\{\alpha = 0, \beta \neq 0\}$ or null $\{\alpha\neq 0, \beta = 0\}$, the classical bootstrap remains valid for the first term in \eqref{equation:decomposition}.

To deal with the second term, \cite{he2023AdaptiveBootstrapTests} proposed a bootstrap statistic inspired by the limiting distribution established in the large-sample asymptotics.  As customary, a superscript $\ast$ denotes a nonparametric bootstrap. In the null case of $(\alpha, \beta) = (0,0)$,  a bootstrap statistic $\mZ_\alpha^\ast\mZ_\beta^\ast$ is constructed as the bootstrap analog of the limit $Z_\alpha Z_\beta$. Since $Z_\alpha$ and $Z_\beta$ represent the limiting distributions of $n^{1/2} \wh\alpha$ and $n^{1/2} \wh\beta$, their bootstrap can defined by their counterparts $\mZ_\alpha^\ast$ and $\mZ_\beta^\ast$, as $n^{1/2} \wh\alpha^\ast$ and $n^{1/2} \wh\beta^\ast$, respectively.

In the AB test, the indicators $I_{\alpha, \lambda_n}$ and $I_{\beta, \lambda_n}$ are replaced, respectively,  with their bootstrap versions: 
\beqrs
I_{\alpha, \lambda_n}^\ast = I(\abs{\wh T_{\alpha}^\ast}\leq\lambda_n, \abs{\wh T_{\alpha}}\leq\lambda_n), \text{ and } \enskip I_{\beta, \lambda_n}^\ast = I(\abs{\wh T_{\beta}^\ast}\leq\lambda_n, \abs{\wh T_{\beta}}\leq\lambda_n),
\eeqrs
where $\wh T_{\alpha}^\ast = n^{1/2} \wh\alpha_{S}^\ast / \wh\sigma_{\alpha}^\ast$ and $\wh T_{\beta}^\ast = n^{1/2} \wh\beta_{M}^\ast / \wh\sigma_{\beta}^\ast$ represent the bootstrap analogues of $\wh T_{\alpha} = n^{1/2} \wh\alpha_{S} / \wh\sigma_{\alpha}$ and $\wh T_{\beta} = n^{1/2} \wh\beta_{M} / \wh\sigma_{\beta}$. When a closed-form expression for $\wh\sigma_{\alpha}^{\ast}$ and $\wh\sigma_{\beta}^{\ast}$ is unavailable, we may estimate them using the sample standard deviations of $\{n^{1/2} \wh\alpha_{S}^{\ast}\}$ and $\{n^{1/2} \wh\beta_{M}^{\ast}\}$, respectively.

This leads to the AB test statistic of the form, 
\beqrs
U^\ast_\tau  =  (\wh\alpha^\ast \wh\beta^\ast-\wh\alpha \wh\beta)\times (1 - I^\ast_{\alpha, \lambda_n}I^\ast_{\beta, \lambda_n})  + \wh\alpha^\ast \wh\beta^\ast \times I^\ast_{\alpha, \lambda_n}I^\ast_{\beta, \lambda_n}.
\eeqrs
If the tuning parameter $\lambda_n$ satisfies $\lambda_n = o(n^{1/2})$, such as $\lambda_n = 2n^{1/2}/\log n$, the bootstrap consistency of $U^\ast_\tau$ has been established by \cite{he2023AdaptiveBootstrapTests}. Empirically, fixing $\lambda_n =2$ has shown satisfactory performance in many simulation experiments. 

{\color{black} The above AB version of the Sobel test has been similarly established for the joint significant test (or MaxP) test in \cite{he2023AdaptiveBootstrapTests}. Along the lines of MaxP test, in more general settings, we can test the mediation effect by testing the independence copula under the vine copula decomposition in (\ref{equation:vine_decomposition}), where $p$-values may be calculated by applying \cite{purkayastha2024FastMIFastConsistent}'s mutual information test. This is worth a further exploration.   

}

\section{R Package \texttt{abima}}\label{section:numerical_studies}
The R package \texttt{abima} \citep{chen2024AbimaAdaptiveBootstrap} provides an implementation of the adaptive bootstrap (AB) test proposed by \cite{he2023AdaptiveBootstrapTests} for linear and generalized linear models. {\color{black} The linear model is a special case of the Gaussian copula SEMs with Gaussian marginals proposed by \cite{hao2023ClassDirectedAcyclic}.} The package offers two primary functions to perform statistical inference for mediation pathways: \textsf{abYlm.Mlm} and \textsf{abYlm.Mglm}. The function \textsf{abYlm.Mlm} performs the AB test for classical linear structural equation models (SEM), while \textsf{abYlm.Mglm} extends the AB test to settings where the outcome follows a generalized linear model. The output of these functions when the exposure $S$ changes from $s^\ast$ to $s$ are shown in Table \ref{tab:mediation_effects}.
\begin{table}[ht]
\centering
\begin{tabular}{|l|l|l|}
\hline
\textbf{Effect} & \textbf{Description} & \textbf{P-value} \\
\hline
\texttt{NIE} (Natural Indirect Effect) & Estimated indirect effect  & \texttt{p\_value\_NIE} \\
\hline
\texttt{NDE} (Natural Direct Effect) & Estimated direct effect & \texttt{p\_value\_NDE} \\
\hline
\texttt{NTE} (Natural Total Effect) & Estimated total effect & \texttt{p\_value\_NTE} \\
\hline
\end{tabular}
\caption{Descriptions of Mediation Effects and Corresponding AB Test $p$-values}
\label{tab:mediation_effects}
\end{table}
We illustrate this R-package via {\color{black}experiments} where we generate data based on the following linear SEM:  
\begin{equation*}
\begin{aligned}
M &= \alpha S + \alpha_I + \alpha_{X,1} X_1 + \alpha_{X,2} X_2 + \epsilon_M, \\  
Y &= \beta M + \beta_I + \beta_{X,1} X_1 + \beta_{X,2} X_2 + \gamma S + \epsilon_Y.
\end{aligned}
\end{equation*}  
In this simulated model, the exposure variable $S\sim \text{Ber}(0.5)$, $(X_1,X_2)^T\sim \text{BVN}(0,0,1,1,0)$ and $(\epsilon_M,\epsilon_Y)\sim \text{BVN}(0,0,\sigma_{\epsilon_M}^2,\sigma_{\epsilon_Y}^2,0)$. The three parameters are set at $(\alpha_I, \alpha_{X,1}, \alpha_{X,2}) = (1,1,1)$, $(\beta_I, \beta_{X,1}, \beta_{X,2}) = (1,1,1)$, $\gamma = 1$, and $\sigma_{\epsilon_M} = \sigma_{\epsilon_Y} = 0.5$ as well as the sample size, $n = 500$. The bootstrap sample size is fixed at $199$ (excluding one bad case from the set of 200 bootstrap replicates) with a nominal level $0.05$.  

\begin{figure}[!htp]
	\centerline{\renewcommand{\arraystretch}{0.8} %
		\begin{tabular}{c}			
\psfig{figure=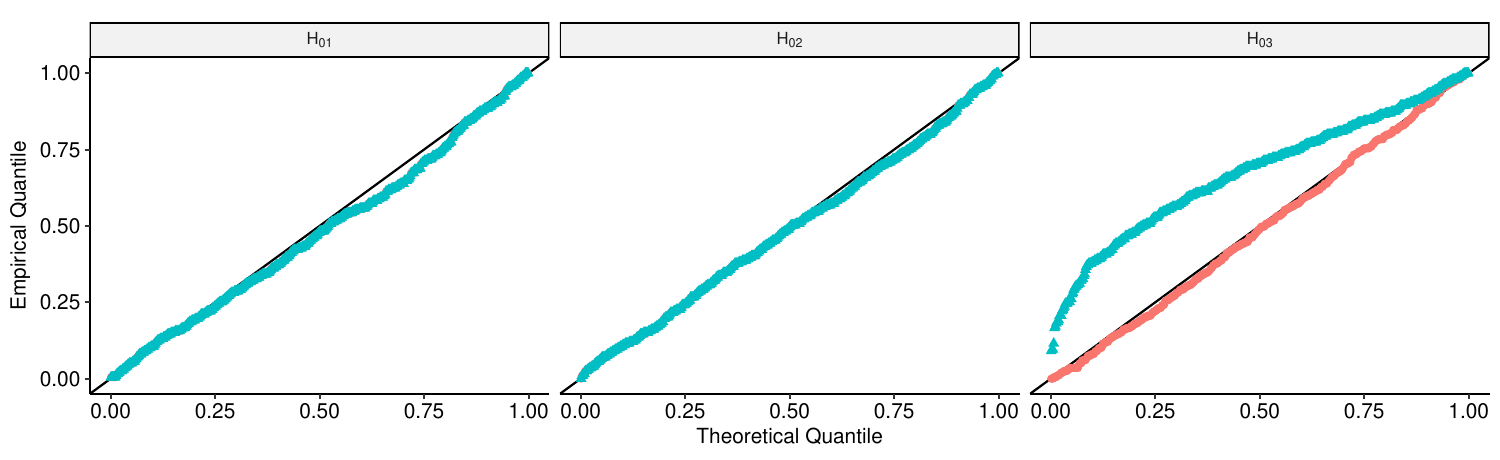,width=6in,angle=0}
		\end{tabular}
	}
	\captionsetup{font = footnotesize}
	\caption{Quantile-Quantile plots of $p$-values under the fixed null hypothesis with $n = 500$ for the classical ($\triangle$) and adaptive ($\circ$) bootstrap methods.
    }
	\label{fig:type_i_error}
\end{figure}

To evaluate type I error control, we simulate data under a fixed null hypothesis across $500$ independent replications and estimate the distribution of $p$-values using the R-package \texttt{abima}. We consider three null hypotheses: $H_{01}\colon (\alpha, \beta) = (0,0.5)$, $H_{02}\colon (\alpha, \beta) = (0.5,0)$, and $H_{03}\colon (\alpha, \beta) = (0,0)$. Quantile-Quantile plots for $n = 500$ are presented in \Cref{fig:type_i_error}. As shown in \Cref{fig:type_i_error}, the classical bootstrap method, despite its widespread use, exhibits conservativeness under the singular null $H_{03}$, whereas the AB test effectively controls type I error across all null scenarios.  

\begin{figure}[!htp]
	\centerline{\renewcommand{\arraystretch}{0.8} %
		\begin{tabular}{cc}			
\psfig{figure=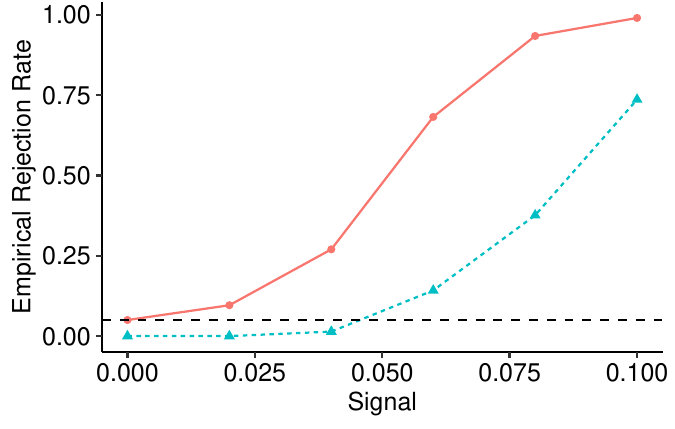,width=3in,angle=0}&\psfig{figure=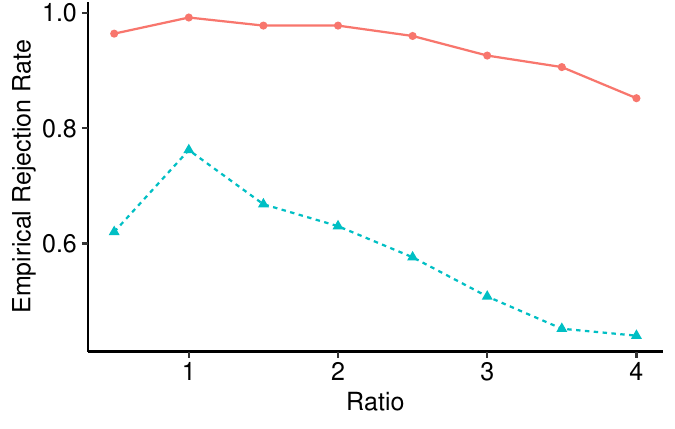,width=3in,angle=0}\\
		\end{tabular}
	}
	\captionsetup{font = footnotesize}
	\caption{Empirical rejection rates (powers) under the fixed alternative hypothesis with $n = 500$ for the classical ($\triangle$) and adaptive ($\circ$) bootstrap methods.
    }
	\label{fig:power}
\end{figure}

We further examine the statistical power under the alternative hypothesis with $500$ independent replications. We consider two settings: (i) fix $\alpha = \beta$ and vary the signal strength $\alpha$ from $0$ to $0.1$; (ii) fix the size of mediation effect $\alpha \beta = 0.01$ and vary the ratio $\alpha/\beta$ from $0.5$ to $4$. The corresponding empirical rejection rates (powers) are shown in \Cref{fig:power}. Due to its conservative nature, the classical bootstrap method exhibits lower power compared to the AB test. 

\section{Concluding Remarks}\label{section:conclusion}
As pointed out in \citet{czado2025VineCopulaBased}, the class of copula structural equation models (CoSEMs), including both Gaussian copula SEMs and vine copula SEMS,  provides a powerful paradigm to analyze mediation pathways beyond what the Gaussian linear SEM can do. We review several recent advancements in copula SEMs with key extensions to accommodate non-Gaussian data of mixed types.  A central focus of the pathway analysis lies on hypothesis testing procedures to assess the presence or absence of certain mediation pathways. To this end, the recent literature has introduced adaptive bootstrap methods that facilitate the identification of different null subspaces, thereby ensuring the validity of hypothesis testing. Many future research directions are worth exploration, including a formal development of causal estimands within the vine copula SEM framework, the refinement of adaptive bootstrap methods for pathway identification in high-dimensional settings with numerous nodes, and the study of spillover effects of mediators that are spatially correlated.  

Inspired by Dr. Czado's strong dedication to software development for vine copulas over her entire career, we built and released an R package \verb\abima\ in the R CRAN to perform adaptive bootstrap tests. There is {\color{black} much room  } for improvement of this software package, including adding a key function \verb\CoSEM\ devoted to analytics presented in this paper for the mediation analysis in the context of the copula structural equation models and beyond. This is worth serious effort in the future.

\bibliographystyle{apalike}
\bibliography{refs}
\end{document}

%% file: definition.tex
\ifnum\draft=1
\usepackage[left=30mm, right=28mm, top=2cm, bottom=2cm]{geometry}

\usepackage{titlesec} %
\titleformat{\section} %
  {\vspace{-0.4 cm}\centering\large\bfseries} %
  {\thesection.} %
  {0.5 em} %
  {\vspace{-0.15 cm}} %

\titleformat{\subsection}
  {\vspace{-0.4 cm}\centering\it} %
  {\thesubsection.} %
  {0.5em} %
  {\vspace{-0.25 cm}} %

\fi

\usepackage{amssymb, amsthm, amsmath, bm, color} %

\usepackage{lineno} %
\usepackage{graphicx} %
\usepackage{multirow} %
\usepackage{booktabs} %
\usepackage{epsfig} %
\usepackage{natbib} %
\usepackage{setspace} %
\usepackage{enumitem} %
\usepackage{threeparttable} %
\usepackage{url} %
\usepackage{doi} %

\usepackage{algorithm, algorithmic} %

\usepackage{hyperref,caption}
\usepackage{cleveref}

\usepackage{tikz}

\numberwithin{equation}{section} %
\graphicspath{{figs/}}

\def\beqr{\begin{eqnarray}}
\def\eeqr{\end{eqnarray}}
\def\beqrs{\begin{eqnarray*}}
\def\eeqrs{\end{eqnarray*}}

\def\wh{\widehat}
\def\wt{\widetilde}
\DeclareFontFamily{U}{mathx}{\hyphenchar\font45}
\DeclareFontShape{U}{mathx}{m}{n}{
      <5> <6> <7> <8> <9> <10>
      <10.95> <12> <14.4> <17.28> <20.74> <24.88>
      mathx10
      }{}
\DeclareSymbolFont{mathx}{U}{mathx}{m}{n}
\DeclareFontSubstitution{U}{mathx}{m}{n}
\DeclareMathAccent{\widebar}{0}{mathx}{"73}

\usepackage{stmaryrd}

\newcommand{\defn}{\stackrel{\mbox{{\tiny def}}}{=}}
\newcommand{\trans}{^{\mbox{\tiny{T}}}}

\newcommand{\diag}{\mathrm{diag}}

\newcommand{\cov}{\mathrm{cov}}

\newcommand{\abs}[1]{\vert#1\vert}

\newcommand{\todistribution}{\xrightarrow{\text{d}}}

\newcommand{\eye}{\mathbf{I}}
\newcommand{\bzeros}{\bm{0}}

\DeclareFontFamily{U}{mathx}{\hyphenchar\font45}
\DeclareFontShape{U}{mathx}{m}{n}{
	<5> <6> <7> <8> <9> <10>
	<10.95> <12> <14.4> <17.28> <20.74> <24.88>
	mathx10
}{}
\DeclareSymbolFont{mathx}{U}{mathx}{m}{n}
\DeclareFontSubstitution{U}{mathx}{m}{n}
\DeclareMathAccent{\widecheck}{0}{mathx}{"71}
\DeclareMathAccent{\wideparen}{0}{mathx}{"75}

\newcommand{\bbeta}{\bm{\beta}}

\newcommand{\bTheta}{\bm{\Theta}}

\newcommand{\bSig}{\bm{\Sigma}}

\newcommand{\W}{\mathbf{W}}

\newcommand{\calN}{\mathcal{N}}

\def\bGam{\bm{\Gamma}}
\def\bfW{\mathbf{W}}
\def\bfbeta{\boldsymbol{\beta}}

\usepackage[all,import]{xy}

\usepackage{xcolor}

\newcommand{\mZ}{\mathbb{Z}}

\usepackage[svgnames]{xcolor}
\usepackage{listings}